\documentclass[11pt, journal]{IEEEtran}

\usepackage[table,xcdraw]{xcolor}
\usepackage{amssymb}
\usepackage{amsmath}
\usepackage{amsthm}
\usepackage{babel}
\usepackage{url}
\usepackage{braket}
\usepackage{algorithm}
\usepackage{algorithmicx}
\usepackage{algpseudocode}

\usepackage{comment}

\usepackage{tikz}
\usepackage{circuitikz}
\usepackage{qcircuit}

\theoremstyle{plain}

\theoremstyle{definition}

\newcommand{\sabre}{\textsf{SABRE}}

\newcommand{\quekno}{\texttt{QUEKNO}}
\newcommand{\queko}{\texttt{QUEKO}}

\newcommand{\swap}[2]{\textsc{swap}({#1},{#2})}

\newcommand{\define}{\ensuremath{\triangleq}}
\newcommand{\la}{\langle}
\newcommand{\ra}{\rangle}

\makeatletter
\newcommand{\xRightarrow}[2][]{\ext@arrow 0359\Rightarrowfill@{#1}{#2}}

\usepackage[normalem]{ulem}

\newcommand{\sqgm}{\textsf{SQGM}}
\newcommand{\toqm}{\textsf{TOQM}}
\newcommand{\nassc}{\textsf{NASSC}}

\usepackage{hyperref}

\usepackage{mathtools}
\usepackage{booktabs}   %% For formal tables:
\usepackage{subcaption} %% For complex figures with subfigures/subcaptions
\usepackage{booktabs}
\usepackage{threeparttable}

\begin{document}

%% Title information
%\title{Single Qubit Gates Matter for Quantum Circuit Transformation}        
%{Single-Qubit Gates Are Important for Reducing Quantum Circuit Depth in Qubit Mapping} 
\title{Single-Qubit Gates Matter for Optimising Quantum Circuit Depth in Qubit Mapping\thanks{$^\star$Accepted to \emph{The 2023 International Conference on Computer-Aided Design} (IEEE/ACM ICCAD'23)}}

\author{
\IEEEauthorblockN{Sanjiang Li}\\
\IEEEauthorblockA{\textit{Centre for Quantum Software \& Information}, 
\textit{University of Technology Sydney},
Sydney, Australia \\
sanjiang.li@uts.edu.au}\\
\and
\IEEEauthorblockN{Ky Dan Nguyen}\\
\IEEEauthorblockA{\textit{School of Computer Science},
\textit{University of Sydney},
Sydney, Australia \\ kngu7458@uni.sydney.edu.au}\\
\and
\IEEEauthorblockN{Zachary Clare}\\
\IEEEauthorblockA{\textit{School of Computer Science},
\textit{University of Technology Sydney},
Sydney, Australia\\
 zachary.clare@student.uts.edu.au}\\
\and
\IEEEauthorblockN{Yuan Feng}\\
\IEEEauthorblockA{\textit{Centre for Quantum Software \& Information}, 
\textit{University of Technology Sydney},
Sydney, Australia \\
yuan.feng@uts.edu.au}
}

\date{}
\maketitle

%\thispagestyle{empty}

%% Abstract
%% Note: \begin{abstract}...\end{abstract} environment must come before \maketitle command
\begin{abstract}
Quantum circuit transformation (QCT, a.k.a. qubit mapping) is a critical step in quantum circuit compilation. Typically, QCT is achieved by finding an appropriate initial mapping and using SWAP gates to route the qubits such that all connectivity constraints are satisfied. The objective of QCT can be to minimise circuit size or depth. Most existing QCT algorithms prioritise minimising circuit size, potentially overlooking the impact of single-qubit gates on circuit depth. In this paper, we first point out that a single SWAP gate insertion can double the circuit depth, and then propose a simple and effective method that takes into account the impact of single-qubit gates on circuit depth. Our method can be combined with many existing QCT algorithms to optimise circuit depth. The Qiskit SABRE algorithm has been widely accepted as the state-of-the-art algorithm for optimising both circuit size and depth. We demonstrate the effectiveness of our method by embedding it in SABRE, showing that it can reduce circuit depth by up to 50\% and 27\% on average on, for instance, Google Sycamore and 117 real quantum circuits from MQTBench.
\end{abstract}

\section{Introduction}
\label{sec:intro}

Current Noisy Intermediate-Scale Quantum (NISQ) devices have strict connectivity constraints that limit the execution of 2-qubit gates (such as CX or CZ) to neighbouring qubits only. This requires a transformation of the ideal circuits before running them on real quantum devices, which is commonly known as Quantum Circuit Transformation (QCT), qubit mapping, or layout synthesis. In this paper, we use the terms \emph{qubit mapping} and \emph{(quantum) circuit transformation} interchangeably to refer to this procedure. QCT is an essential component of quantum circuit compilation and has gained widespread interest in areas such as quantum computing \cite{ChildsSU19-qct,Cowtan+19-tket,Nannicini+21_bipmapping,Saeedi+11_synthesis,Venturelli+18_Planner}, electronic design automation \cite{Ash-Saki+19_qure,Deng0L20_codar,Itoko+19_commutation,TanC21-gate_absorption,Xie21dac_commutativity,ZhouFL20_MCTS_iccad,Zhou+20_SAHS,Zulehner+18_Astar}, and computer architecture \cite{Li+19-sabre,Liu+22_not_all_swap,Murali+19,TannuQ19, Zhang+21-time}.

Over the past few years, numerous QCT algorithms have been developed to transform ideal quantum circuits into circuits that can be executed on a specific quantum device with connectivity constraints. The input to these algorithms includes an architecture graph that specifies the connectivity constraints of the targeted quantum device, and an ideal quantum circuit that contains only single- and 2-qubit gates defined in the basic gate library of the device. QCT algorithms typically achieve this transformation by first applying an initial mapping, followed by repeatedly adding SWAP gates to schedule 2-qubit gates for execution on the target device. The objective of QCT can be to minimise circuit size or depth, both in turn can increase the overall fidelity of the transformed circuit. The transformation cost of such a transformation depends on the optimisation objective, which can be measured either by the number of SWAP gates inserted or the difference in circuit depth before and after transformation. Due to the NP-complete nature of the qubit mapping problem \cite{Siraichi+18,TanC21-queko}, exact algorithms can only handle--within reasonable time--circuits with up to 10 qubits, and therefore, most QCT algorithms are heuristic or approximate in nature.

The error rates of 2-qubit gates in present NISQ devices are often 10 times higher than those of single-qubit gates. Consequently, numerous QCT algorithms \cite{Zulehner+18_Astar,Li+19-sabre,LiZF21_fidls,Zhou+20_SAHS,ZhouFL20_MCTS_iccad} prioritise minimising the number of 2-qubit gates as their primary objective, which is essentially determined by the total number of SWAP gates inserted. In many cases, QCT algorithms execute a gate as early as possible and single-qubit gates are often not considered, see, e.g., \cite{Li+19-sabre,ChildsSU19-qct,Cowtan+19-tket,LiZF21_fidls}. In fact, FiDLS \cite{LiZF21_fidls} and an initial implementation of \sabre\ \cite{Li+19-sabre} even completely remove single-qubit gates from the circuit before transformation. 

In the past several years, we have seen the size of quantum computer increases from 5 to 433  qubits\footnote{{https://newsroom.ibm.com/2022-11-09-IBM-Unveils-400-Qubit-Plus-Quantum-Processor-and-Next-Generation-IBM-Quantum-System-Two}}, but qubit coherence time in NISQ devices remains very short. This implies that we can only run a very limited number of quantum operations on each qubit; in other words, we cannot extract meaningful information from very deep quantum circuits on NISQ devices. Thus, minimising the depth of transformed circuits is perhaps a more important objective. There are several QCT algorithms targeting circuit depth, see, e.g., Qiskit's \texttt{StochasticSWAP} and \cite{ChildsSU19-qct,tcad/LaoSAA22}, which is often achieved by encouraging parallel SWAPs or minimising the depth of inserted SWAP circuits. Tan and Cong \cite{TanC20_iccad_optimal} and Zhang et al. \cite{Zhang+21-time} propose exact QCT algorithms for minimising circuit depth. Again, these algorithms can transform only small circuits. Both exact algorithms are also relaxed to obtain approximate algorithms, which could tackle circuits with more qubits than their exact version while still can obtain much better depth results than heuristic-based algorithms like \sabre. However, based on SMT or $A^*$ search, the two approximate algorithms are still not scalable to circuits with 50 or more qubits (see experiments reported in Sec.~\ref{sec:tomq}).

First described in \cite{Li+19-sabre}, \sabre\ was later implemented in Qiskit and is now its default transpiler. As a randomised algorithm \sabre\ can  be used for both circuit size and depth optimisations: we need only run it multiple times and select the best size or depth result. In addition, \sabre\ introduces a `decay' factor to discourage applying SWAP gates on a qubit which was recently swapped. Extensive evaluation on several quantum devices shows that \sabre\ significantly outperforms many state-of-the-art QCT algorithms in both circuit size and depth \cite{li23quekno}. 

Despite this outstanding performance in depth optimisation, the impact of single-qubit gates is overlooked. 
This is because \sabre\ executes a gate whenever it is allowed to do so. In particular, it greedily executes every single-qubit gate if its predecessor has been executed. This sometimes results in unnecessarily much deeper transformed circuits (cf. the example in Fig.s~\ref{fig:sabrerun1} and~\ref{fig:sabrerun2}).

In this paper, we first show  by an example that single-qubit gates are also important and \textbf{a single SWAP gate insertion may double the circuit depth}, and then propose a method that takes into account the impact of single-qubit gates on circuit depth. The idea is to record the \emph{transformation progress} of each physical qubit, hold single-qubit gates until we meet a new 2-qubit gate after them, and introduce a `delay' component based on the qubit progress to discourage those SWAPs that have progressed too much on their qubits. Our method can be combined with many existing QCT algorithms for optimising circuit depth. We demonstrate the effectiveness of our method by embedding it in \texttt{SabreSWAP}---the Qiskit implementation of \sabre's routing process, showing that it can reduce circuit depth by up to 50\% and 27\% on  average on, for instance, Google Sycamore (54Q) and 117 real quantum circuits from MQTBench \cite{quetschlich2022mqtbench}. Similar results are also observed on IBM Q Tokyo (20Q) and Rochester (53Q) and synthesised \quekno\ benchmarks for evaluating depth optimality  \cite{li23quekno}, where the average improvements over \sabre\ are 15\% and 17\%, respectively.

It was found that the relaxed {\toqm} algorithm \cite{Zhang+21-time} performs very well on 20-qubit IBM Q Tokyo and could beat \sabre\ in terms of circuit depth by on average 20\%. This paper will also compare our algorithm and {\toqm} on larger quantum devices. Results (reported in Sec.~\ref{sec:tomq}) show that the relaxed version of {\toqm} is not yet scalable to larger devices such as the 54Q Google Sycamore, while \sabre\ and our algorithm, called \sqgm\ (for \textbf{S}ingle-\textbf{Q}ubit \textbf{G}ates \textbf{M}atter), can process within a few seconds. 

Another recent work \cite{Liu+22_not_all_swap} combines qubit mapping with 2-qubit block re-synthesis and commutation-based gate cancellation. Their algorithm, called \nassc, enriches \sabre\ with the above optimisation techniques. In this paper, we also empirically compare \nassc\ with our  \sqgm. Results show that, on MQTBench circuits and 54Q Sycamore, \sqgm\ outperforms \nassc\ by 12\% in terms of circuit depth; and, if we apply a post-routing commutative gate cancellation (which \nassc\ has already included) to our algorithm, then \sqgm\ outperforms \nassc\ by 17\%.

The remainder of this paper is organised as follows. We recall relevant backgrounds about quantum circuits and quantum circuit transformation as well as \sabre\ in Sec.~\ref{sec:prelimaries}, and present our method in Sec.~\ref{sec:sqgm}. We then evaluate and compare our method with \sabre, {\toqm} \cite{Zhang+21-time}, {\nassc} \cite{Liu+22_not_all_swap} in Sec.~\ref{sec:evaluation}. The last section concludes the paper.

\section{Preliminaries}\label{sec:prelimaries}

This section first recalls some relevant background in quantum computing  and then introduces the preliminaries of quantum circuit transformation. In the end of this section, we recall the QCT algorithm \sabre.

\subsection{Quantum Circuits}
Quantum algorithms are commonly described using quantum circuits, which are analogous to classical combinational circuits. A quantum circuit comprises a sequence of quantum gates that act on qubits (quantum bits). Quantum gates are unitary transformations. An $n$-qubit gate is represented as a $2^n\times 2^n$ unitary matrix. 
Some well-known single-qubit gates include:
\begin{equation*}
\resizebox{\hsize}{!}{%
$X = \begin{pmatrix}
    0& 1\\ 1 & 0
    \end{pmatrix}$,\quad 
    $H = \frac{1}{\sqrt{2}}\begin{pmatrix}
    1& 1\\ 1 & -1
    \end{pmatrix}$,\quad  
    $S = \begin{pmatrix}
    1& 0\\ 0 & i
    \end{pmatrix}$, \quad 
    $T = \begin{pmatrix}
    1& 0\\ 0 & e^{i\frac{\pi}{4}}
    \end{pmatrix}.$
    }
\end{equation*}
CX (also called CNOT) and CZ are two most common 2-qubit gates. For any computational basis state $\ket{i}\ket{j}$, CX and CZ map $\ket{i}\ket{j}$ to, respectively, $\ket{i}\ket{i\oplus j}$ and $(-1)^{i\cdot j}\ket{i}\ket{j}$, where $\oplus$ denotes exclusive-or and $\cdot$ denotes logical conjunction. 

Any quantum gate can be implemented using, i.e., \emph{decomposed} into, single-qubit and CX gates. Furthermore, we can approximate any quantum gate to arbitrary accuracy using the $H$, $S$, $T$, and CX gates. The SWAP gate, which swaps the states of two qubits, can be implemented using three CX gates. That is, $\swap{p}{q} = \textsc{CX}(p,q)\textsc{CX}(q,p)\textsc{CX}(p,q)$.

Although different quantum devices may have different universal sets of quantum gates, the 2-qubit gate in these sets is usually either CX or CZ. Since the functionality of a single-qubit gate is not (directly) relevant in quantum circuit transformation, a gate acting on qubit $q_i$ is simply denoted as $\la q_i\ra$, while a CX or CZ gate with control qubit $q_i$ and target qubit $q_j$ is denoted as $\la q_i,q_j\ra$.

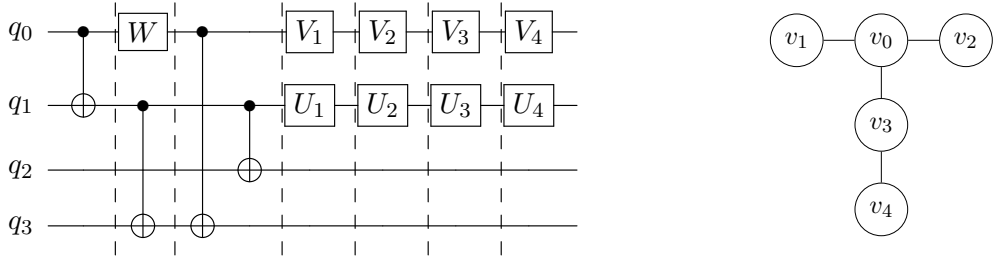
\begin{figure*}
\begin{tabular}{ll}
\begin{tabular}{l}
\hspace{-5em}
\centerline{
\Qcircuit @C=0.8em @R=1.1em {
\lstick{q_{0}}  & \ctrl{1} \barrier{3} &\gate{W}  \barrier{3} &\ctrl{3} &\qw  \barrier{3}\ & \gate{V_1} \barrier{3} &\gate{V_2} \barrier{3} & \gate{V_3} \barrier{3} & \gate{V_4}  & \qw\\
\lstick{q_{1}}  &\targ & \ctrl{2} &\qw &\ctrl{1} & \gate{U_1} & \gate{U_2}& \gate{U_3}& \gate{U_4}  & \qw  \\
\lstick{q_{2}} &\qw   & \qw &\qw &\targ & \qw&\qw  & \qw& \qw   & \qw\\
\lstick{q_{3}} &\qw   & \targ &\targ &\qw & \qw & \qw& \qw  & \qw &\qw   \\
  }
  }
\end{tabular}
& 
\hspace*{-40mm}
\scalebox{0.9}{
\begin{tabular}{l}
\begin{tikzpicture}[auto,node distance=3cm]
    \node[shape=circle,draw=black] (v0) at (1.25,2.5) {$v_0$};
    \node[shape=circle,draw=black] (v1) at (0,2.5) {$v_1$};
    \node[shape=circle,draw=black] (v2) at (2.5,2.5) {$v_2$};
    \node[shape=circle,draw=black] (v3) at (1.25,1.25) {$v_3$};
    \node[shape=circle,draw=black] (v4) at (1.25,0) {$v_4$};
    \path (v0) edge (v1);
    \path (v0) edge (v2);
    \path (v0) edge (v3);
    \path (v3) edge (v4);
\end{tikzpicture}
\end{tabular}
}
\\
\end{tabular}
\caption{A logical circuit with depth 7 (left) and the architecture graph of IBM Q Ourense (right)}
\label{fig:logical_circuit}
\end{figure*}

A circuit $C$ is usually given as a sequence of gates ($g_0,g_1,\ldots,g_{m-1}$), but this does not mean that the $(i+s)$-th gate should be executed after the $i$-th gate for all $i \geq 0, s \geq 1$ with $0<i+s<m$. In fact, two gates can be executed in parallel if they do not act on a common qubit. Naturally, we partition $C$ into layers while putting each gate as front as possible (so that it can be executed  
at the earliest time). The number of layers is called the \emph{depth} of the circuit. For example, the circuit in Fig.~\ref{fig:logical_circuit} can be represented as
\begin{equation*}
\begin{split}
     C\! =\! \big(\la q_0,\! q_1\ra, \la q_0\ra, \la q_1,\!q_3\ra, \la q_0,\! q_3\ra, \la q_1,\!q_2\ra,\\ 
     \la q_0\ra, \la q_0\ra, \la q_0\ra, \la q_0\ra,  \la q_1\ra, \la q_1\ra, \la q_1\ra, \la q_1\ra \big).
\end{split}
\end{equation*}
The circuit has 7 layers and thus a depth of 7. Specifically, its third layer contains two gates, viz. $\la q_0,\! q_3\ra, \la q_1,\! q_2\ra$. 

This paper represents a quantum device as an undirected graph $G = (V, E)$, called the \emph{architecture graph} of the quantum device, where $V$ denotes the set of physical qubits and $E$ the set of permitted 2-qubit interactions. In other words,  $(v,v')$ is an edge in $E$ iff a 2-qubit gate acting on qubits $v, v'$ is executable on the device. As $G$ is an undirected graph, $(v,v')\in E$ iff $(v',v)\in E$. Fig.~\ref{fig:logical_circuit} (right) shows the architecture graph of an artificial device.

\subsection{Quantum Circuit Transformation}
\label{sec:qct}

In the quantum circuit model, it is common for the algorithm designer to not have a targeted quantum device in mind when developing the algorithm.
Let $C$ be an ideal circuit representing a quantum algorithm and $G=(V,E)$ the architecture graph of the target quantum device. We assume gates in $C$ have been decomposed into elementary gates supported by the physical device and our task is to transform $C$ into a functionally equivalent circuit where every 2-qubit gate acts on two neighbouring nodes in $G$.

In the following, we refer to $C$ as a \emph{logical} circuit and call the transformed circuit a \emph{physical} circuit. The qubits in the logical (physical) circuit are called logical (physical) qubits. It is worth noting that, in the context of QCT, the use of the term ``logical'' should not be confused with its usage in error correction. Device-supported single-qubit gates can be executed directly if its predecessor in the circuit has been executed. A 2-qubit gate is \emph{directly executable} on $G$ if its two qubits are neighbours in $G$ and its predecessors in the circuit have been executed.

The circuit transformation task involves two steps: \emph{initial mapping} and \emph{qubit routing}. 
A typical QCT algorithm runs as follows: Select an initial mapping $\tau_0$. After removing all or a subset of executable gates under $\tau_0$ from $C$, repeat the following two procedures alternatively until there are no gates left: (i) transform the mapping into a new mapping by inserting one or several SWAP gates and (ii) remove all or a subset of executable gates.

\vspace*{2mm}
\noindent\textbf{Notation:} Throughout this paper, we will use $p,q,p',q'$, possibly with subscripts, to denote logical qubits, and use $u,v,u',v'$, possibly with subscripts, to denote physical qubits, i.e., vertices in the architecture graph.

\begin{figure*}
\centering
\begin{tabular}{r}
\centerline{
\Qcircuit @C=0.8em @R=1.1em{
\lstick{q_{0}\mapsto v_{1}} & \ctrl{1}  &\gate{W}  &\qw &\qw  &\qw
       & \qw  &\qw &\ctrl{1} & \targ & \ctrl{1} &\qw & \qw &\qw   & \qw &\qw & \qw \\
\lstick{q_{1}\mapsto v_{0}}  &\targ & \ctrl{2}  &\ctrl{1} & \gate{U_1}  & \gate{U_2}& \gate{U_3}  & \gate{U_4}  & \targ &\ctrl{-1} 
&\targ & \ctrl{2} & \gate{V_1} &\gate{V_2} & \gate{V_3} & \gate{V_4} & \qw
\\
\lstick{q_{2}\mapsto v_{2}} &\qw   &\qw  &\targ &\qw & \qw &\qw & \qw  & \qw& \qw  & \qw& \qw & \qw  & \qw &\qw & \qw& \qw \\
\lstick{q_{3}\mapsto v_{3}} &\qw  & \targ &\qw & \qw &\qw  &\qw  & \qw &\qw & \qw& \qw&\targ & \qw& \qw  & \qw &\qw & \qw  
\gategroup{1}{9}{2}{11}{.7em}{--} 
}
}
\end{tabular}
\caption{A SABRE transformation with depth 15.}\label{fig:sabrerun1}
\end{figure*}
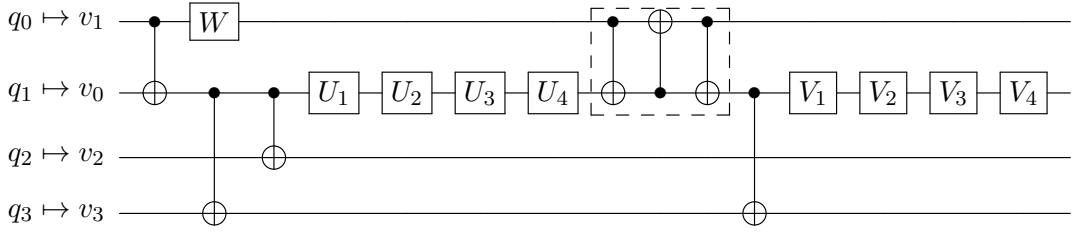

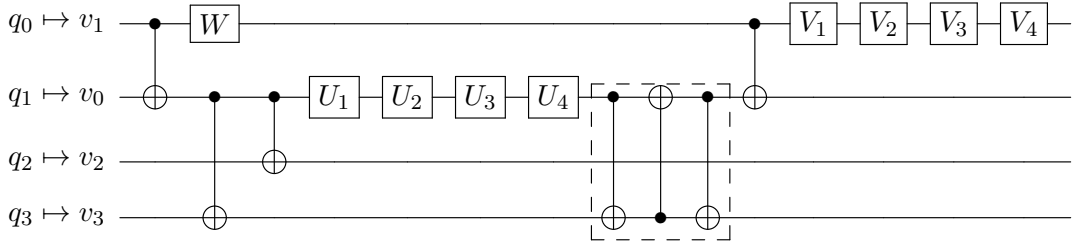
\begin{figure*}
\centering
\begin{tabular}{r}
\hspace{2em}

\centerline{
\Qcircuit @C=0.8em @R=1.1em {
\lstick{q_{0}\mapsto v_{1}} & \ctrl{1}  &\gate{W}  &\qw &\qw  &\qw 
       & \qw   &\qw &\qw &\qw &\qw &\ctrl{1} & \gate{V_1} &\gate{V_2} & \gate{V_3} & \gate{V_4} & \qw \\
\lstick{q_{1}\mapsto v_{0}}  &\targ & \ctrl{2} &\ctrl{1} & \gate{U_1}  & \gate{U_2} & \gate{U_3}  & \gate{U_4}    & \ctrl{2} & \targ & \ctrl{2} & \targ &\qw   &\qw   & \qw &\qw & \qw \\
\lstick{q_{2}\mapsto v_{2}} &\qw   & \qw &\targ &\qw & \qw &\qw & \qw  & \qw & \qw  & \qw& \qw &\qw   &\qw   & \qw &\qw & \qw\\
\lstick{q_{3}\mapsto v_{3}} &\qw   & \targ &\qw &\qw& \qw &\qw &\qw &\targ &\ctrl{-2} &\targ & \qw & \qw& \qw  & \qw &\qw & \qw  
\gategroup{2}{9}{4}{11}{.7em}{--}
  }
  }
\end{tabular}
\caption{Another SABRE transformation with depth 15.}\label{fig:sabrerun2}
\end{figure*}

\subsection{\sabre}\label{sec:sabre}

We first recall the heuristics of \sabre\ \cite{Li+19-sabre}. Let $S$ be a set of 2-qubit gates and $\pi: P \to V$ the current mapping, where $P$ and $V$ denote the set of logical qubits in a given circuit $C$ and physical qubits in a given architecture graph $G$. Define 
\begin{align}
V(\pi,S) \define \sum_{ g\in S} \text{dist}\big(\pi(g.{p_0}),\pi(g.{p_1})\big),
\end{align}
where $g.{p_0}$ and $g.{p_1}$ are the first and second qubits that gate $g$ acts on, and $\text{dist}(v,v')$ is the length of a shortest path from $v$ to $v'$ in the architecture graph. That is, $V(\pi,S)$ measures the total distances between the qubit pair of all 2-qubit gates in $S$, under the mapping $\pi$. The smaller $V(\pi,S)$ is the better $\pi$ is.  

Let $F$ be the 2-qubit gates in the current front layer and $X$ the set of 2-qubit gates in the extended set (which are 2-qubit gates that follow those in the front layer but could  occupy on several subsequent layers), which are used to `look ahead' and increase the applicability of a candidate mapping to `future' gates. For each edge $(v_i,v_j)$ in the architecture graph, we define three heuristics:
\begin{align}
H_\textsf{basic} &= V(\pi',F)\\
H_\textsf{lookahead} &= \frac{1}{|F|}V(\pi',F)+w\cdot \frac{1}{|X|}V(\pi',X) \label{eq:sabre_h_lookahead}
\end{align}
\begin{align}
 H_\textsf{decay} = \max\big(\text{decay}(v_i),\text{decay}(v_j)\big)\cdot H_\textsf{lookahead}, \label{eq:sabre_h_decay}
\end{align}
where $\pi'\define \pi_{i,j}\circ \pi$, $\pi_{i,j}$ is the permutation obtained by swapping $v_i,v_j$, $\text{decay}(v_i)\geq 1$ is a \emph{decay factor} and $w\in [0,1]$ is a weight. The initial decay factor of a qubit is 1, and is incremented by some pre-defined \emph{decay rate} $\delta$ every time that qubit is involved in a SWAP, thus encouraging the algorithm to insert SWAP gates acting on different qubits, i.e., they can be executed in parallel. In the search procedure, \sabre\ inserts $\swap{v_i}{v_j}$ if it has the minimum $H$ value.

The key factor driving \sabre's success is its unique approach to integrating the initial mapping and routing processes. Initially, a random mapping is applied, and then the routing process is run using either the ``lookahead'' or ``decay'' search strategies until all gates in the circuit $C=(g_1,\ldots,g_m)$ have been executed. The final mapping is then used as the initial mapping for transforming the reverse circuit $C^{-1}=(g_m,\ldots,g_1)$, which is routed using the same approach. This forward-backward transformation may iterate multiple times, and the final mapping obtained is used as the actual initial mapping to transform $C$. As a result, the algorithm takes into account the global information of the circuit when generating the initial mapping.

\section{Our Method}\label{sec:sqgm}
In this section we describe our method. We start with a motivating example and then describe in detail our approach to enhance a QCT algorithm like \sabre.

\subsection{A Motivating Example}
Consider the logical circuit as well as the architecture graph of IBM Q Ourense shown on Fig.~\ref{fig:logical_circuit}, which will be the running example in this paper. Fig.~\ref{fig:sabrerun1} and Fig.~\ref{fig:sabrerun2} show two examples of a \sabre\ transformation, where the initial mapping $\tau$ is defined as $q_0\mapsto v_1, q_1\mapsto v_0, q_2\mapsto v_2, q_3\mapsto v_3$. Since $\tau(q_1)=v_0$, the CX gates $CX(q_0,q_1), CX(q_1,q_3), CX(q_1,q_2)$ are all executable. We transform these gates immediately. Since single-qubit gates are always executable, we transform $W$ and $U_1$ to $U_4$ immediately. As $\tau(q_0)=v_1$ and $\tau(q_3)=v_3$ are not neighbours in $G$, $CX(q_0,q_3)$ is not executable. Note that $CX(q_0,q_3)$ is the only CX gate in the current front layer and there are no CX gates in the extended set. We have two SWAP gate candidates, viz. $\swap{v_0}{v_1}$ and $\swap{v_0}{v_3}$, each corresponding to an edge in $G$. In general, a candidate SWAP gate $\swap{v}{v'}$ corresponds to an edge $(v,v')$ in $G$ such that either $v$ or $v'$ is in the image of $\tau$ and either $\tau^{-1}(v)$ or $\tau^{-1}(v')$ is in $F$, where $\tau^{-1}$ is the inverse of $\tau$. Suppose we use the ``lookahead'' heuristics. By Eq.~\ref{eq:sabre_h_lookahead}, it is easy to see that both $\swap{v_0}{v_1}$ and $\swap{v_0}{v_3}$ have score $H=1$ and we can insert either to transform the circuit, resulting in the physical circuits depicted in Fig.s~\ref{fig:sabrerun1} and~\ref{fig:sabrerun2} respectively. After the insertion, all the remaining gates are executable and the transformation is completed. 

\begin{figure*}
\centering
\begin{tabular}{r}
\hspace{2em}
\centerline{
\Qcircuit @C=0.8em @R=1.1em {
\lstick{q_{0}\mapsto v_{1}} & \ctrl{1}  &\gate{W}  &\qw &\ctrl{1} & \targ & \ctrl{1}  & \gate{U_1}  & \gate{U_2}& \gate{U_3}  & \gate{U_4} 
&\qw  &\qw   \\
\lstick{q_{1}\mapsto v_{0}}  &\targ & \ctrl{2}  &\ctrl{1} & \targ &\ctrl{-1} 
&\targ & \ctrl{2} & \gate{V_1} &\gate{V_2} & \gate{V_3} & \gate{V_4} & \qw
\\
\lstick{q_{2}\mapsto v_{2}} &\qw   &\qw  &\targ &\qw & \qw &\qw & \qw  & \qw& \qw  & \qw& \qw & \qw  \\
\lstick{q_{3}\mapsto v_{3}} &\qw  & \targ &\qw & \qw &\qw  &\qw  &\targ & \qw &\qw & \qw& \qw & \qw  
\gategroup{1}{5}{2}{7}{.7em}{--}
  }
  }
\end{tabular}
\caption{A transformation with depth 11.}\label{fig:betterrun}
\end{figure*}
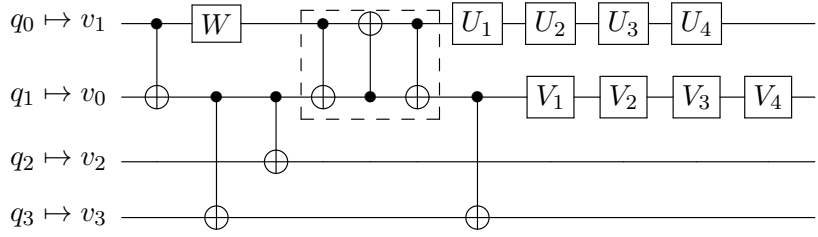

These two transformations show that sometimes \textbf{a single SWAP insertion can double the circuit depth}! Actually, suppose we replace the  single-qubit gate lists $V_1,\ldots, V_4$ and $U_1,\ldots,U_4$ in  Fig.~\ref{fig:logical_circuit} with  $V_1,\ldots, V_k$ and $U_1,\ldots,U_k$ for $k\in \mathbb{N}$. Then a single SWAP insertion as shown in Fig.~\ref{fig:sabrerun1} or Fig.~\ref{fig:sabrerun2} increases the circuit depth from $k+3$ to $2k+7$!  This, however, can be avoided by, for example, inserting the same SWAP gate $\swap{v_0}{v_1}$ before the single-qubit gate $U_1$, resulting a better transformation with depth $k+7$ (see Fig.~\ref{fig:betterrun}).

The example shown above suggests that single-qubit gates can significantly affect the depth of the transformed circuit. In the following, we show how the intuition illustrated in Fig.~\ref{fig:betterrun} can be developed into a general method for improving the depth optimality of QCT algorithms.

\subsection{Qubit Progress}
To track the transformation progress of a quantum (logical) circuit $C$ on a target architecture $G$, each physical qubit $v$ on $G$ is associated with a progress indicator denoted as $PG(v)$. 

Initially, all $PG(v)$ values are set as 0. Whenever a single-qubit gate $U(p)$ acting on logical qubit $p$ is transformed, we increment the progress on the corresponding physical qubit $v=\tau(p)$ by 1, where $\tau$ is the current mapping. If a 2-qubit gate (CX or CZ) $V(p,p')$ acting on logical qubits $p$ and $p'$ is transformed, the progress on the physical qubits $v=\tau(p)$ and $v'=\tau(p')$ should be \textbf{realigned}, by setting both $PG(v)$ and $PG(v')$ to $\max(PG(v),PG(v'))+1$. When a SWAP gate is inserted on $v=\tau(p)$ and $v'=\tau(p')$, the progress on the physical qubits $v$ and $v'$ should also be realigned, by setting both $PG(v)$ and $PG(v')$ to $\max(PG(v),PG(v'))+3$, considering that each SWAP gate is decomposed into three consecutive CX gates. 

With the transformation progress indicator $PG$, it is possible to identify which qubit is progressing too quickly and which is lagging behind. For example, suppose we apply the transformation depicted in Fig.~\ref{fig:sabrerun1} and have transformed $CX(q_0,q_1)$, $CX(q_1,q_3)$, $CX(q_1,q_2)$, and single-qubit gates $W$ and all $U_i$ for $1\leq i\leq 4$. Then, the progress on the qubits are as follows: $PG(v_0)=7$, $PG(v_1)=2$, $PG(v_2)=3$, and $PG(v_3)=2$, see Table~\ref{tab:pg}. Apparently, $v_0$ is progressing too rapidly.
\begin{table}[h]
\centering
\caption{The qubit progress immediately after a particular gate has been transformed as in Fig.~\ref{fig:sabrerun1}. }\label{tab:pg}
\resizebox{0.45\textwidth}{!}{%
\begin{tabular}{c|c|c|c|c|c}
 & $CX(q_0,q_1)$ & $W$ & $CX(q_1,q_3)$ & $CX(q_1,q_2)$ & $U_4$  \\ \hline
$PG(v_0)$ & 1 & 1 & 2 & 3 & 7 \\ 
$PG(v_1)$  & 1 & 2 & 2 & 2 & 2 \\ 
$PG(v_2)$  & 0 & 0 & 0 & 3 & 3 \\ 
$PG(v_3)$  & 0 & 0 & 2 & 2 & 2 
\end{tabular}
}
\end{table}

\subsection{Single-qubit Gate Buffer}\label{sec:buffer}
 In contrast to \sabre\ and many other QCT algorithms, when a single-qubit gate is next in line to be transformed on physical qubit $v$, it is not transformed immediately. Instead, it is placed in a \emph{buffer} (a temporary storage), which, denoted as $S(v)$, is an ordered set. The search for the next executable 2-qubit gate continues. When a new executable 2-qubit gate $CX(p,p')$ is encountered, all single-qubit gates in $S(v)$ and $S(v')$, where $v=\tau(p)$ and $v'=\tau(p')$, are transformed one by one. If no such executable 2-qubit gate is found, these single-qubit gates in $S(v)$ and $S(v')$ remain reserved until the next SWAP gate is selected for insertion. For example, in Fig.~\ref{fig:sabrerun1}, before inserting the SWAP gate, we store the executable single-qubit gates $W$ and $U_i, 1 \leq i \leq 4$ in, respectively, $S(v_1)$ and $S(v_0)$. Meanwhile, $S(v_2)=S(v_3)=\varnothing$. Since these single-qubit gates are reserved, the qubit progress at this time is $PG(v_0)=PG(v_2)=3, PG(v_1)=1, PG(v_3)=2$.  

\subsection{SWAP Selection and Insertion}
We can enhance the heuristic $H$ adopted by a QCT algorithm  with the qubit progress indicator $PG$. Take \sabre\ as an example. For each SWAP on some edge $(v,v')$ in $G = (V, E)$, let $\nu(v,v')=\max(PG(v),PG(v'))$. Then, the enhanced heuristic score associated with that SWAP is \begin{equation}\label{eq:hsqgm}
    H_\textsf{sqgm} = H + \frac{\nu(v,v')}{|V|},
\end{equation} where $H$ is the score computed by Eq.~\ref{eq:sabre_h_lookahead} or Eq.~\ref{eq:sabre_h_decay}. Continuing with our running example, suppose that all gates executable by the initial mapping have been processed (i.e., executed and removed or stored in their buffers). Recall that we have two candidate SWAPs, i.e., those on edges $(v_0,v_1)$ and $(v_0,v_3)$. Recall also that the front layer contains one gate $CX(q_0,q_3)$ and the extended set is empty. By Eq.~\ref{eq:sabre_h_lookahead}, the lookahead scores of these edges are $H(v_0,v_1)=H(v_0,v_3)=1$. Thus, \sabre\ will select one SWAP from $(v_0,v_1)$ or $(v_0,v_3)$. Observe that $\nu(v_0,v_1) = \nu(v_0,v_3) = 3$ (cf.~Sec.~\ref{sec:buffer}). Thus, the enhanced heuristic scores for both $(v_0,v_1)$ or $(v_0,v_3)$ are $1 + 3/5$ (since IBM Q Ourense has 5 qubits). Hence, the enhanced \sabre\ also picks randomly from $(v_0,v_1)$ and $(v_0,v_3)$.

Suppose the selected best SWAP operation acts on physical qubits $v=\tau(p), v'=\tau(p')$. We need to decide where to insert this SWAP. This depends on the qubit progress and the buffer sizes of $v,v'$. If $PG(v)=PG(v')$, then we insert the SWAP gate in the front of $v,v'$, i.e., before the first single-qubit gates on $v,v'$. If $PG(v) < PG(v')$, then we insert the SWAP gate in front of the $k$-th single-qubit gate in $S(v)$ (after transforming them one by one), where $k=\min(PG(v')-PG(v), |S(v)|)$. The case when $PG(v) > PG(v')$ is analogous. After swapping, we need to exchange the buffers of $v,v'$. 

In our example, suppose we picked $(v_0,v_1)$ as the SWAP gate. Since $PG(v_0)=3>1=PG(v_1)$ and $S(v_1)=(W), S(v_0)=(U_1,\ldots,U_4)$, we first transform $W$ (and remove it from $S(v_1)$) and then insert $\swap{v_0}{v_1}$ immediately after $CX(q_1,q_2)$. Meanwhile, we  swap the buffers for $v_1$ and $v_0$. That is, after the swap insertion, we have $PG(v_0)=PG(v_1)=6$ and $S(v_1)=(U_1,\ldots,U_4)$ and $S(v_0)=S(v_2)=S(v_3)=\varnothing$. This is precisely the transformation described in Fig.~\ref{fig:betterrun}. Note that after swap insertion, the remaining gates are executable and we also transform all single-qubit gates in the buffer if there are any.

Algorithm~\ref{alg:sqgm} presents the pseudocode of our algorithm \sqgm. The procedure \textsc{getSwapCands}$(F, G)$ on line 7 obtains SWAP candidates from the physical neighbours of the qubits involved in the front layer $F$. For every non-executable gate $g(p_0, p_1) \in F$,\footnote{Note that $g$ is guaranteed to be a 2-qubit gate as all single-qubit gates before $g$ are either executed or stored in a buffer.} we retrieve the corresponding physical qubits $v_0 = \tau(p_0), v_1 = \tau(p_1)$ and construct $N_0, N_1$ via the target architecture graph $G = (V, E)$, where we define $N_i = \{(v_i, v): v \in V, (v_i, v) \in E\}$, i.e., the set of edges incident to $v_i$ in $G$. Then, the SWAP candidates are precisely $N_0 \cup N_1$. The procedure \textsc{addResolvedSuccessors}$(F, g)$ on line 37 checks every successor gate $h$ of $g$, and adds $h$ to $F$ if \emph{all} of $h$'s predecessor gates have been executed.

\begin{algorithm}
\caption{Single-Qubit Gates Matter}
\label{alg:sqgm}
\small
\begin{algorithmic}[1]
\Require{An ideal circuit $C$, a target architecture graph $G = (V, E)$, an initial mapping $\tau_0$}
\Ensure{A transformation of $C$ satisfying $G$'s connectivity constraints}
\State Initialise a buffer $S(v_i)$ for each $v_i \in V$
\State $F \gets \text{front layer of } C$
\State $\tau \gets \tau_0$
\While{$F \neq \varnothing$}
    \State $\textsc{executableGateList} \gets \text{all}\ g \in F\ \text{executable on}\ G$
    \If{$\textsc{executableGateList} = \varnothing$}
       \State $\textsc{swapCands} \gets \textsc{getSwapCands}(F, G)$
        \State Compute $H_\textsf{sqgm}$ for each $\textsc{swap} \in \textsc{swapCands}$
        \State $(p_0, p_1) \gets \text{a}\ \textsc{swap} \text{ with minimum } H_\textsf{sqgm}$
        \If{$PG(\tau(p_0)) < PG(\tau(p_1))$}
           \State $v_0, v_1 \gets \tau(p_0), \tau(p_1)$ 
        \Else
           \State $v_0, v_1 \gets \tau(p_1), \tau(p_0)$
    \EndIf
    \State $k \gets \min(PG(v_1) - PG(v_0), |S(v_0)|)$
    \For{$g_0$ in the front $k$ gates of $S(v_0)$}
        \State Execute and remove $g_0$ from $S(v_0)$
        \State $PG(v_0) \gets PG(v_0) + 1$
    \EndFor
    \State $S(v_0), S(v_1) \gets S(v_1), S(v_0)$
    \State $PG(v_{0}) \gets  PG(v_{1}) \gets \max(PG(v_0), PG(v_1)) + 3$
    \State Update $\tau$ with $(p, q)$
    \Else    
        \For{$g \in \textsc{executableGateList}$}
            \If{$g(p)$ is a single-qubit gate}
                \State Add $g$ to $S(\tau(p))$ \textit{\# Do not execute $g$ yet}
            \ElsIf{$g(p_0, p_1)$ is a 2-qubit gate}
                \For{$i \in \{0,1\}$ and $g_i \in S(\tau(p_i))$}
                    \State Execute and remove $g_i$ from $S(\tau(p_i))$
                    \State $PG(\tau(p_{i})) \gets PG(\tau(p_{i})) + 1$
                \EndFor
                \State Execute $g$
                \State $PG(\tau(p_{0})) \gets PG(\tau(p_{1})) \gets$
                \State $\,\,\,\,\,\,\,\,\,\max(PG(\tau(p_0)),PG(\tau(p_1))) + 1$
            \EndIf
            \State Remove $g$ from $F$
            \State $\textsc{addResolvedSuccessors}(F, g)$
        \EndFor
        \State Clear \textsc{executeGateList}
    \EndIf
\EndWhile
\end{algorithmic}
\end{algorithm}

\section{Experiments and Evaluation}\label{sec:evaluation}
Our algorithm (implemented in Python 3) and benchmarks as well as experimental results are available %on the authors' GitHub repository.
at \url{https://github.com/ebony72/sqgm}.
All our experiments were run on a computer with AMD Ryzen 5 5600 CPU, 32 GB RAM and AMD Radeon RX 6600 GPU.

\sabre\ has recently been assembled in Qiskit. In this paper, we choose this Qiskit (version 0.39.4) implementation of \sabre\ and adopt the advanced `decay' heuristic, where the weight $w$ and the decay rate $\delta$ are 0.5 and 0.001 as usual (cf. Sec.~\ref{sec:sabre}). When comparing our algorithm with \sabre, or comparing different versions of \sabre, we take the same initial mapping generated from the current \texttt{SabreLayout} module (the initial mapping pass of \sabre), but use different routing modules. 

The current Qiskit \texttt{SabreSWAP} (the routing pass of \sabre) uses internal accelerators written in Rust which are not open-source. Therefore, when implementing our \sqgm\ routing module, we go back to Qiskit version 0.33.0 and modify the \texttt{SabreSWAP} module there. The 0.39.4 version of \texttt{SabreSWAP} (henceforth \sabre39) performs slightly (around 2-10\%) better than the 0.33.0 version (henceforth \sabre33), thanks to several optimisations. Unless otherwise specified, the \sabre\ algorithm always refers to \sabre39. 

\subsection{Benchmarks and Architecture Graphs}\label{sec:bench}
\begin{figure}
\centering
\begin{tabular}{c}
    \includegraphics[width=0.48\textwidth]{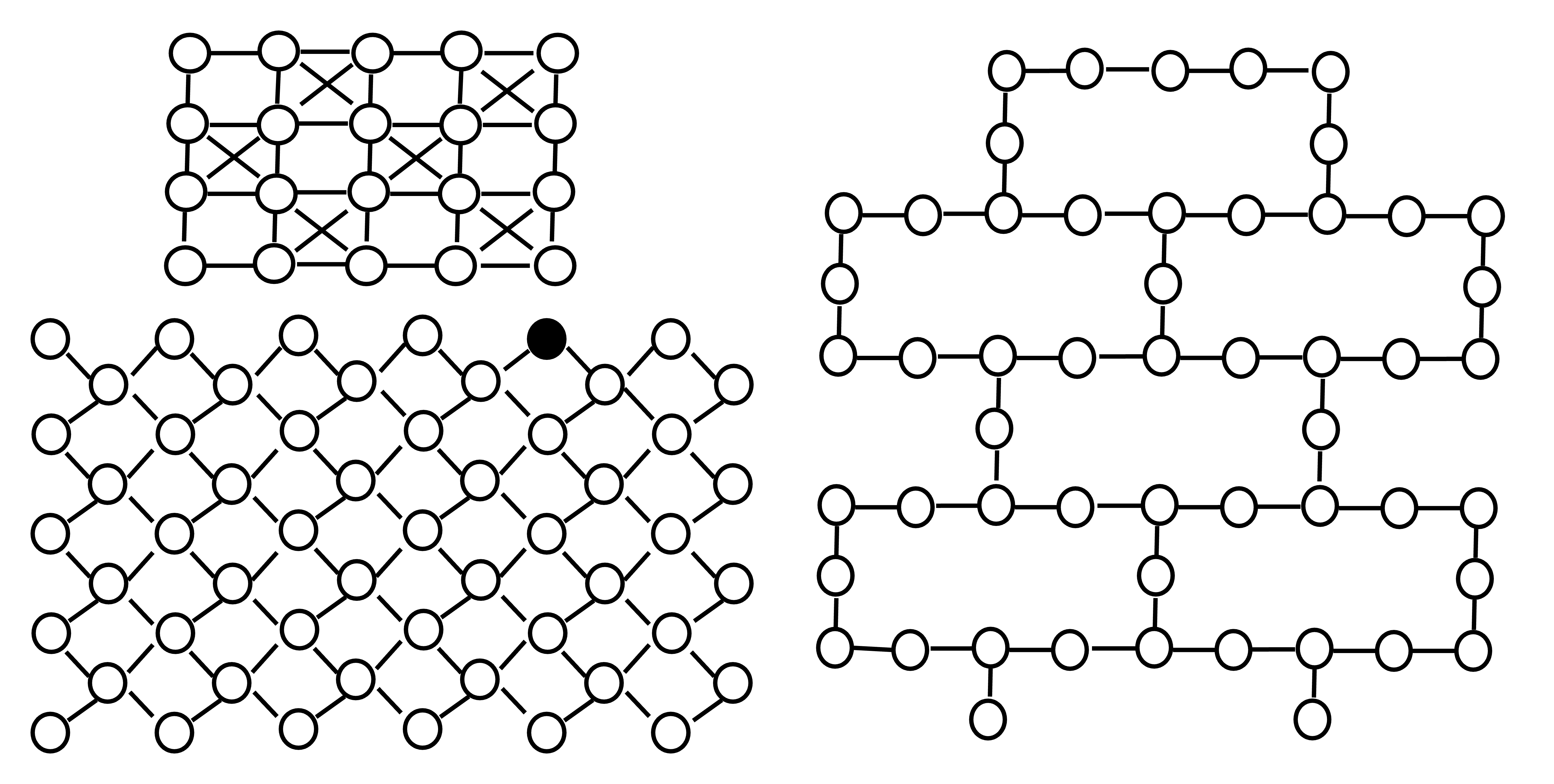} 
\end{tabular}
\caption{The architecture graphs of IBM Q Tokyo (20Q, upper left), Google Sycamore (54Q, bottom left) and IBM Q Rochester (53Q, right), where the black node in Google Sycamore denotes a bad qubit node.} \label{fig:sycamore&rochester}
\end{figure}

To compare the depth optimality performance of our algorithm with \sabre\ and two other state-of-the-art algorithms, we considered extensive synthetic and real benchmarks on three architecture graphs: IBM Q Tokyo (20Q), IBM Q Rochester (53Q), and Google Sycamore (53Q and 54Q), see Fig.~\ref{fig:sycamore&rochester}. Note that Google Sycamore has a bad qubit node, i.e., the black node in Fig.~\ref{fig:sycamore&rochester} (bottom left). Sometimes we will remove this node as well as its incident edges. We call this modified architecture the 53Q Sycamore. 

The \queko\ benchmark \cite{TanC21-queko} is designed to evaluate the depth optimality of QCT algorithms for various architectures and has been adopted by several researchers in evaluating QCT algorithms \cite{TanC20_iccad_optimal,Zhang+21-time,li23quekno}. 
By design, each \queko\ circuit has a zero-cost optimal transformation, i.e., if we can find the right initial mapping, then no SWAP is required to transform the circuit. Our evaluation reveals that, on IBM Q Tokyo and \queko\ circuits, \sabre\  can often find an optimal transformation within 100 repeats. So, in the following, we only consider 54Q \queko\ benchmarks, for which an optimal transformation is difficult to find due to the large search space.

This shortcoming of \queko\ has been addressed in \cite{li23quekno}, where \quekno\ benchmarks are proposed for evaluating the optimality of QCT algorithms. Each \quekno\ circuit has a nonzero optimal transformation cost which can be closely estimated. For each architecture, we evaluate \sqgm\ and compare algorithms on the corresponding \quekno\ benchmark set, which contains 120 circuits that all have a known \emph{near-optimal} transformation cost.  

Besides these synthesised benchmarks, we also extracted real quantum circuits from \href{https://www.cda.cit.tum.de/mqtbench/}{MQTBench} \cite{quetschlich2022mqtbench}, to be run on 54Q Google Sycamore. The circuit set we selected include 117 circuits, which are obtained by selecting
\begin{itemize}
    \item All scalable benchmarks;
    \item Qubit Range: between 50 and 54;
    \item Target-independent Level: Qiskit as the used compiler; and
    \item Target-dependent Native Gates Level: targeted native gate-set from IBM, Qiskit as the used compiler, and optimisation level Opt. 3.
\end{itemize}
These circuits include well-known quantum circuits, such as Amplitude Estimation, Graph State, GHZ State, Grover's, QAOA, QFT, Quantum Phase Estimation, Quantum Walk, VQE, W-State. Before sending these circuits to a QCT algorithm, we decompose each non-standard gate into standard IBM basic gates. Table~\ref{tab:mqtbench} shows name and depth of all those circuits with 53 qubits.

\subsection{Compare \sabre\ and \sqgm}
We first compare the performances of \sqgm\ and \sabre\ on several large benchmark sets. Recall that \sabre\ can often find the optimal solution for \queko\ benchmarks on IBM Q Tokyo. To make a meaningful comparison, we evaluate \sqgm\ on \quekno\ benchmarks \cite{li23quekno} and the three architecture graphs in Fig.~\ref{fig:sycamore&rochester}. For each architecture, the corresponding \quekno\ benchmark set contains 120 circuits.  We summarise the results in Table~\ref{tab:summary_quekno}, where we can see the improvement on 20Q Tokyo is about 15\% and the improvements on 53Q Rochester and 53Q Sycamore are 17\% and 23\%. Note that here we used 53Q Sycamore instead of the 54Q version because the corresponding \quekno\ benchmarks were designed for 53Q Sycamore, for which we have known near-optimal depth costs.

\begin{table}[h]
\centering
\caption{Compare \sabre\ with \sqgm\ on \quekno\ benchmarks \cite{li23quekno} for evaluating depth optimality. }\label{tab:summary_quekno}
\scalebox{0.75}{
\begin{tabular}{c|c|c|c}
benchmark sets & 20Q\_depth\_Tokyo & 53Q\_depth\_Rochester & 53Q\_depth\_Sycamore  \\ 
device & IBM Q Tokyo & IBM Q Rochester & 53Q Google Sycamore \\ \hline
\sabre & 2.177 & 2.748 & 2.577\\
\sqgm & 1.854 & 2.289  & 1.992 \\ 
ratio & \textbf{0.852} & \textbf{0.833} & \textbf{0.773}\\
\bottomrule
\end{tabular}
}
\begin{tablenotes}
\item [a] Note: each entry in row `\sabre' is the geomean of 120 results, each of which is obtained as the best depth ratio (i.e., the ratio of the smallest depth of \sabre\ transformed circuit among 5 repeats and that of the input circuit); similar interpretation applies to entries in row `\sqgm'; entries in the `ratio' row denote the ratios of the corresponding geomeans for \sqgm\ and \sabre.
\end{tablenotes}
\end{table}

Since the above benchmarks are all synthesised, we also compared \sabre\ and \sqgm\  on the 117 MQT benchmarks \cite{quetschlich2022mqtbench} and 54Q Google Sycamore. Table~\ref{tab:mqtbench} shows the results for all 53Q circuits, from which we can see that the average improvement of \sqgm\ over \sabre\ is 27\% and the best improvement could be as high as 50\%.

\begin{table*}[t!]
\caption{Comparison of \sqgm\ and \nassc\ against \sabre\ on 53-qubit MQT circuits \cite{quetschlich2022mqtbench} and Google Sycamore (54Q), where `+CC' indicates that the post-routing \texttt{Commutative Gate Cancellation} pass has been applied on the respective algorithm (cf.~Sec.~\ref{sec:nassc}). The depth columns show the input depths or the depths of the  circuits transformed by an algorithm; the ratio columns show the ratios of the depths of an algorithm and those  of \sabre. }\label{tab:mqtbench}
\centering
\resizebox{0.95\textwidth}{!}{%
\begin{tabular}{ll|l|ll|ll|ll|ll}
\multicolumn{2}{c|}{Circuit}& \multicolumn{1}{c|}{\sabre} & \multicolumn{2}{c|}{\sabre+CC} & \multicolumn{2}{c|}{\sqgm} &
\multicolumn{2}{c|}{\sqgm+CC} & \multicolumn{2}{c}{\nassc}\\ 
        name & depth & depth & depth & ratio & depth & ratio & depth & ratio & depth & ratio \\ \hline\hline
ae\_indep\_qiskit\_53.qasm                               & 614  & 1961          & 1736          & 0.89          & 1489          & 0.76          & 1412  & 0.72          & 1756  & 0.90          \\
ae\_nativegates\_ibm\_qiskit\_opt3\_53.qasm & 872 & {1845} & {1876} & {1.02} & {1718} & 0.93          & 1722  & 0.93          & 1666  & 0.90          \\
dj\_indep\_qiskit\_53.qasm                               & 55   & 168           & 166           & 0.99          & 128           & 0.76          & 132   & 0.79          & 172   & 1.02          \\
dj\_nativegates\_ibm\_qiskit\_opt3\_53.qasm              & 58   & 180           & 176           & 0.98          & 135           & 0.75          & 132   & 0.73          & 171   & 0.95          \\
ghz\_indep\_qiskit\_53.qasm                              & 54   & 157           & 149           & 0.95          & 79            & 0.50          & 75    & 0.48          & 149   & 0.95          \\
ghz\_nativegates\_ibm\_qiskit\_opt3\_53.qasm             & 56   & 126           & 137           & 1.09          & 81            & 0.64          & 75    & 0.60          & 134   & 1.06          \\
graphstate\_indep\_qiskit\_53.qasm                       & 21   & 43            & 45            & 1.05          & 39            & 0.91          & 36    & 0.84          & 38    & 0.88          \\
graphstate\_nativegates\_ibm\_qiskit\_opt3\_53.qasm      & 21   & 35            & 30            & 0.86          & 29            & 0.83          & 29    & 0.83          & 23    & 0.66          \\
qft\_indep\_qiskit\_53.qasm                              & 419  & 1859          & 1489          & 0.80          & 1282          & 0.69          & 1174  & 0.63          & 1459  & 0.78          \\
qft\_nativegates\_ibm\_qiskit\_opt3\_53.qasm             & 370  & 1491          & 1315          & 0.88          & 1312          & 0.88          & 1217  & 0.82          & 1412  & 0.95          \\
qftentangled\_indep\_qiskit\_53.qasm                     & 421  & 1967          & 1811          & 0.92          & 1367          & 0.69          & 1245  & 0.63          & 1720  & 0.87          \\
qftentangled\_nativegates\_ibm\_qiskit\_opt3\_53.qasm    & 373  & 1569          & 1493          & 0.95          & 1305          & 0.83          & 1145  & 0.73          & 1373  & 0.88          \\
qpeexact\_indep\_qiskit\_53.qasm                         & 619  & 2236          & 1877          & 0.84          & 1521          & 0.68          & 1447  & 0.65          & 1931  & 0.86          \\
qpeexact\_nativegates\_ibm\_qiskit\_opt3\_53.qasm        & 523  & 1743          & 1611          & 0.92          & 1404          & 0.81          & 1427  & 0.82          & 1527  & 0.88          \\
qpeinexact\_indep\_qiskit\_53.qasm                       & 619  & 2384          & 1934          & 0.81          & 1608          & 0.67          & 1485  & 0.62          & 1952  & 0.82          \\
qwalk-v-chain\_indep\_qiskit\_53.qasm                    & 29447 & 42033         & 41931         & 1.00          & 35973         & 0.86          & 34872 & 0.83          & 40850 & 0.97          \\
realamprandom\_indep\_qiskit\_53.qasm                    & 214  & 2724          & 2475          & 0.91          & 1693          & 0.62          & 1529  & 0.56          & 2016  & 0.74          \\
realamprandom\_nativegates\_ibm\_qiskit\_opt3\_53.qasm   & 226  & 2605          & 2466          & 0.95          & 1671          & 0.64          & 1482  & 0.57          & 1965  & 0.75          \\
su2random\_indep\_qiskit\_53.qasm                        & 214  & 2587          & 2243          & 0.87          & 1619          & 0.63          & 1423  & 0.55          & 1942  & 0.75          \\
su2random\_nativegates\_ibm\_qiskit\_opt3\_53.qasm       & 226  & 2667          & 2329          & 0.87          & 1674          & 0.63          & 1555  & 0.58          & 1774  & 0.67          \\
twolocalrandom\_indep\_qiskit\_53.qasm                   & 214  & 2635          & 2290          & 0.87          & 1628          & 0.62          & 1541  & 0.58          & 1851  & 0.70          \\
twolocalrandom\_nativegates\_ibm\_qiskit\_opt3\_53.qasm  & 226  & 2643          & 2571          & 0.97          & 1635          & 0.62          & 1457  & 0.55          & 2019  & 0.76          \\
wstate\_indep\_qiskit\_53.qasm                           & 160  & 200           & 208           & 1.04          & 178           & 0.89          & 170   & 0.85          & 180   & 0.90          \\
wstate\_nativegates\_ibm\_qiskit\_opt3\_53.qasm          & 213  & 259           & 258           & 1.00          & 235           & 0.91          & 226   & 0.87          & 153   & 0.59          \\ \hline\hline
\textbf{geomean}                                         &      &               &               & \textbf{0.93} &               & \textbf{0.73} &       & \textbf{0.69} &       & \textbf{0.83}
\end{tabular}
}
\end{table*}

\vspace*{2mm}

\noindent \textbf{Role of the Heuristic}\quad 
The above evaluation shows that \sqgm\ can significantly improve the depth optimality of \sabre. An experiment on 54Q Sycamore and ten \queko\ benchmarks `54QBT\_25CYC\_QSE' shows that, if we use $H_\textsf{decay}$ (Eq.~\ref{eq:sabre_h_decay}) instead of $H_\textsf{sqgm}$ (Eq.~\ref{eq:hsqgm}), the improvement is around 9\%, instead of 33\%  obtained from $H_\textsf{sqgm}$. This suggests that the \sqgm\ heuristic (for selecting the right SWAP candidate) plays a more important role than simply adjusting the position of the selected SWAP  by comparing the qubit progress. 

\vspace*{2mm}

\noindent \textbf{Time Complexity}\quad \sqgm\ has the same time complexity as \sabre, i.e., $O(m\cdot n^{2.5})$, where $m$ is the number of gates and $n$ the number of qubits in the quantum device.
Using internal accelerators, \sabre39 is faster than \sabre33 and \sqgm\ in practice. On the 117 MQT benchmarks, the runtimes of \sqgm\ ranges from 0.09 to 28.01 seconds and,  on average, \sqgm\ takes 1.5x and 0.5x more time than  \sabre39 and \sabre33.

\subsection{Performance Under Different Repeats} 
\sabre\ and \sqgm\ are randomised algorithms. By design, better results can be obtained if we run them more times. In above, we have seen that \sqgm\ performs significantly better than \sabre, but the conclusion was obtained for a fixed repeat number, viz. 5. What may happen if we run both \sabre\ and \sqgm, say, 50 times: will \sqgm\ still perform  significantly better? This subsection is devoted to examining their performance under different repeat numbers.

\begin{figure}
\centering
\begin{tabular}{l}
\includegraphics[width=0.4\textwidth]{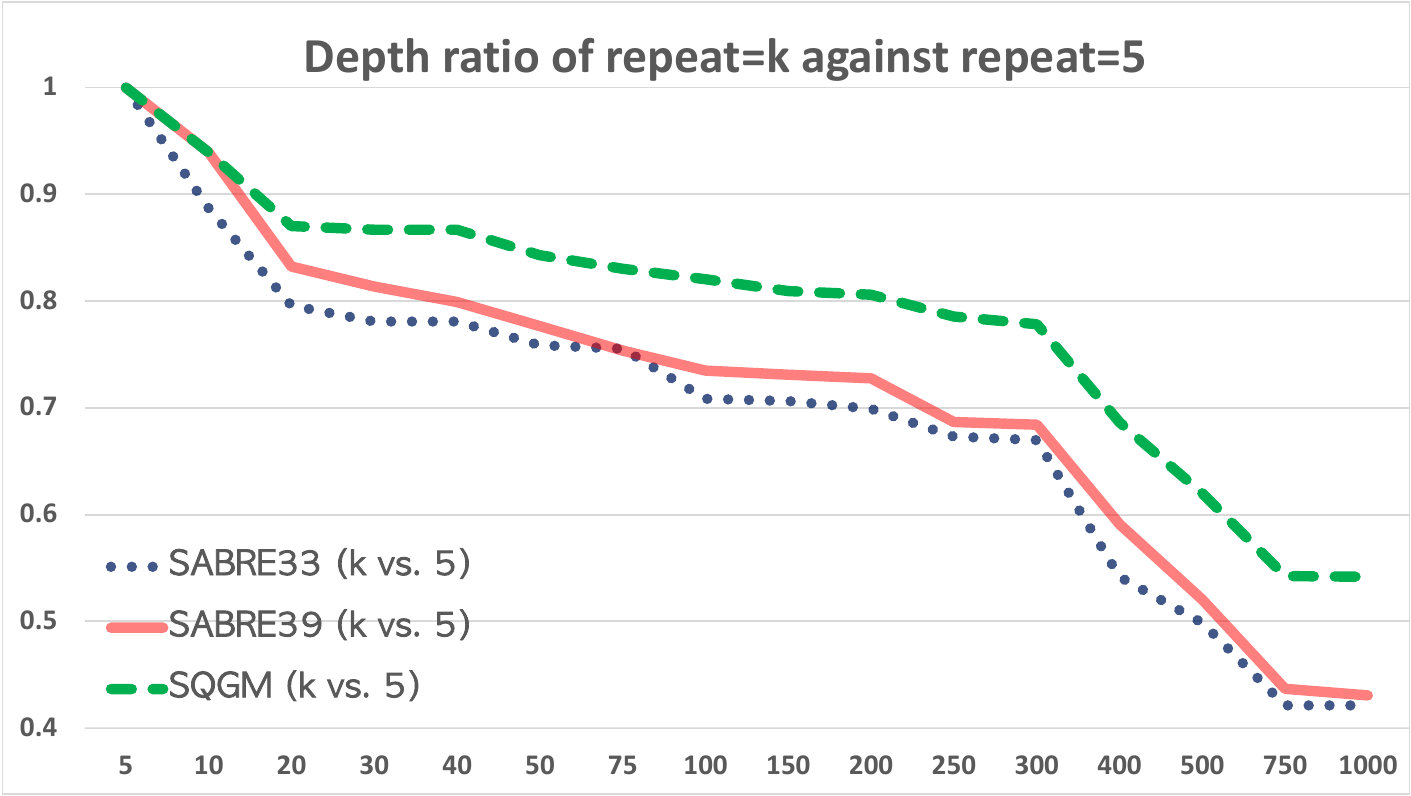}\\
\includegraphics[width=0.4\textwidth]{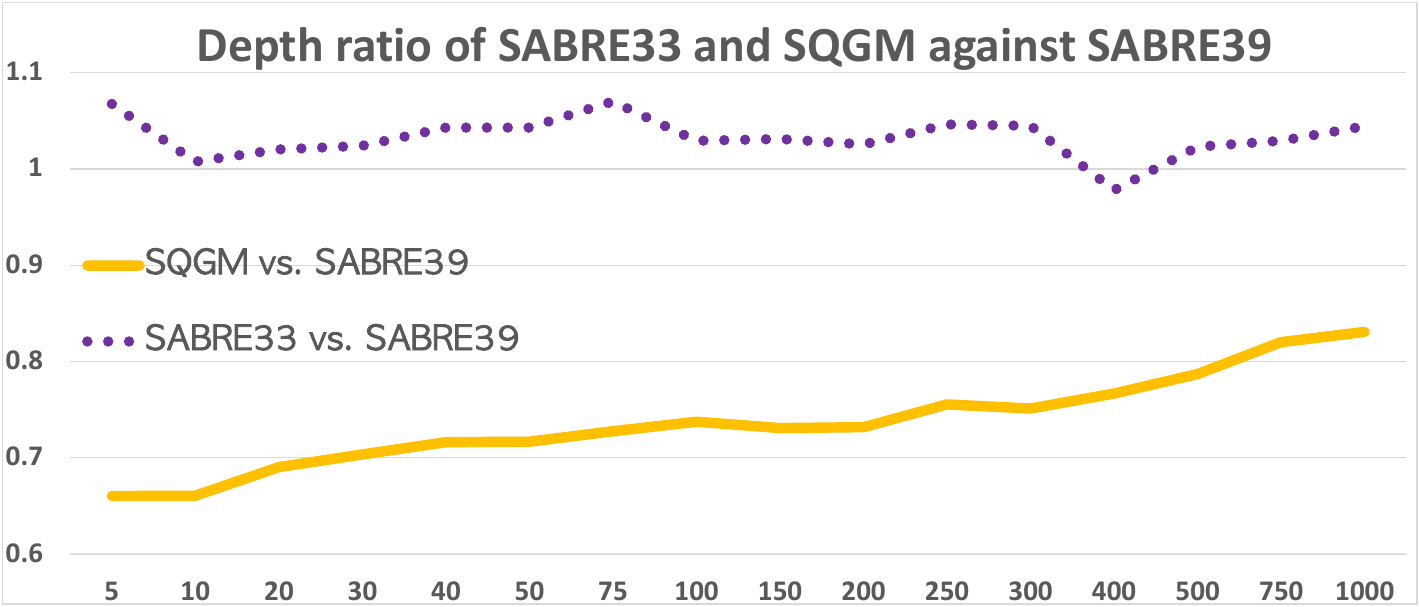}\\
\end{tabular} 
\caption{Changes of $\Delta^k$ (Eq.~\ref{eq:Delta^k}) for \sabre33, \sabre39, and \sqgm\ (top) and changes of $\Lambda^k$ (Eq.~\ref{eq:Lambda^k}) of \sabre33 and \sqgm\ against \sabre39 (bottom) on the 10 \queko\ `54QBT\_25CYC\_QSE' benchmarks and 54Q Sycamore. 
}
\label{fig:repeat}
\end{figure}

We compare the performance of \sabre33, \sabre39, and \sqgm\  on the ten \queko\ `54QBT\_25CYC\_QSE' benchmarks under different repeats, ranging from 5 to 1000. For each algorithm and each repeat number $k$, let $\delta^k_i$ be the minimum depth after $k$ repeats for circuit $i$. Let also
\begin{align}\label{eq:Delta^k}
    \Delta^k \equiv  \text{the geomean of $\delta^k_i / \delta^5_i$ for $0\leq i\leq 9$}.
\end{align}
Then, $1-\Delta^k$ quantifies the average improvement when the repeat number is increased from 5 to $k$. Fig.~\ref{fig:repeat} (top) shows the change of $\Delta^k$ for \sabre33, \sabre39 and \sqgm\ with the repeat number $k$ ranging from 5 to 1000. We note that the scale in the $x$-axis of this figure is not uniform. It is clear that \sabre33 and \sabre39 have the same trends and the depth of the transformed circuit can be reduced on average by 15\%, 20\%, 25\%, 30\%, 40\%, 55\% if we repeat, respectively, 20, 40, 75, 250, 400, 750 times. A similar pattern is also observed for \sqgm, though its circuit depth reduction decreases less rapidly than \sabre33 and \sabre39 with increasing repeat number.

In addition, we compare the performance of \sabre33 (\sqgm) against that of \sabre39 on circuit $i$ by setting $\lambda_i^k$ as the ratio between the minimum depths after $k$ repeats for \sabre33 (\sqgm) and \sabre39. Let
\begin{align}\label{eq:Lambda^k}
\Lambda^k \equiv \text{the geomean of $\lambda_i^k$ for $0\leq i\leq 9$}.
\end{align}
Then $1-\Lambda^k$ quantifies the average improvement of \sabre33 (\sqgm) against \sabre39. The result in Fig.~\ref{fig:repeat} (bottom) shows that \sabre39\ often outperforms \sabre33 and the improvement of \sqgm\ against \sabre\ decreases with $k$ from 35\% to 15\%. It suggests in particular that \sqgm\ still outperforms \sabre\ significantly even after 1000 repetitions.

A similar pattern was also observed on the ten \quekno\ `53QBT\_depth\_Sycamore\_large\_opt\_10\_2.55' benchmarks.

\subsection{Compare with \toqm}
\label{sec:tomq}

\begin{table*}[htb]
\caption{Comparison of \sabre\ and \sqgm\ with {\toqm} on 20Q IBM Q Tokyo and 10 \quekno\ benchmarks ``20QBT\_depth\_Tokyo\_large\_opt\_10\_2.55''.}
\label{tab:toqm_on_tokyo}
\centering
\resizebox{0.8\textwidth}{!}{%
\begin{tabular}{llll|ll|ll|ll|ll|ll}
\multicolumn{4}{c|}{Circuit Information}       & \multicolumn{2}{c|}{{\toqm}}                          & \multicolumn{2}{c|}{\sabre\ repeat=5} & \multicolumn{2}{c|}{\sqgm\ repeat=5} & \multicolumn{2}{c|}{\sabre\ repeat=100} & \multicolumn{2}{c}{\sqgm\ repeat=100} \\
No.   & \#gate    & \#CX      & depth     & depth     & ratio                                 & depth        & ratio               & depth         & ratio            & depth         & ratio                & depth          & ratio             \\ \hline\hline
0 & 760                        & 229                        & 93                        & 188                        & 2.02                        & 266                        & 2.86                        & 195                        & 2.10                        & 232                        & 2.49 & 180 & 1.94 \\
1 & 769 & 229 & 97 & 178 & 1.84 & 220 & 2.27 & 164 & 1.69 & 193 & 1.99 & 157 & 1.62 \\
2 & 725                        & 236                        & 93                        & 176                        & 1.89                        & 245                        & 2.63                        & 229                        & 2.46                        & 209                        & 2.25 & 173 & 1.86 \\
3 & 756                        & 227                        & 90                        & 179                        & 1.99                        & 253                        & 2.81                        & 202                        & 2.24                        & 222                        & 2.47 & 171 & 1.90 \\
4 & 850                        & 254                        & 99                        & 188                        & 1.90                        & 284                        & 2.87                        & 229                        & 2.31                        & 248                        & 2.51 & 200 & 2.02 \\
5 & 795                        & 258                        & 95                        & 191                        & 2.01                        & 262                        & 2.76                        & 208                        & 2.19                        & 244                        & 2.57 & 187 & 1.97 \\
6 & 864                        & 274                        & 110                       & 192                        & 1.75                        & 291                        & 2.65                        & 215                        & 1.95                        & 256                        & 2.33 & 215 & 1.95 \\
7 & 801                        & 245                        & 95                        & 178                        & 1.87                        & 284                        & 2.99                        & 216                        & 2.27                        & 250                        & 2.63 & 204 & 2.15 \\
8 & 893                        & 280                        & 109                       & 188                        & 1.72                        & 311                        & 2.85                        & 206                        & 1.89                        & 235                        & 2.16 & 191 & 1.75 \\
9 & 837                        & 238                        & 101                       & 188                        & 1.86                        & 318                        & 3.15                        & 247                        & 2.45                        & 259                        & 2.56 & 201 & 1.99 \\
\hline\hline
\multicolumn{4}{c|}{\textbf{geomean}}  & \textbf{} & \textbf{1.88} &              & \textbf{2.77}       & \textbf{}     & \textbf{2.14}    &               & \textbf{2.39}        & \textbf{}      & \textbf{1.91}    
\end{tabular}%
}
\end{table*}

\begin{table*}[htb]
\caption{Comparison of \sabre\ and \sqgm\ with {\toqm} on 54Q Sycamore and 10 \quekno\ benchmarks ``20QBT\_depth\_Tokyo\_large\_opt\_10\_2.55''.}
\label{tab:toqm_on_sycamore}
\centering
\resizebox{0.8\textwidth}{!}{%
\begin{tabular}{llll|ll|ll|ll|ll|ll}
\multicolumn{4}{c|}{Circuit Information}       & \multicolumn{2}{c|}{{\toqm}}                          & \multicolumn{2}{c|}{\sabre\ repeat=5} & \multicolumn{2}{c|}{\sqgm\ repeat=5} & \multicolumn{2}{c|}{\sabre\ repeat=100} & \multicolumn{2}{c}{\sqgm\ repeat=100} \\
No.   & \#gate    & \#CX      & depth     & depth     & ratio                                 & depth        & ratio               & depth         & ratio            & depth         & ratio                & depth          & ratio             \\\hline\hline
0 & 760                        & 229                        & 93                        & 274                        & 2.95                        & 311                        & 3.34                        & 202                        & 2.17                        & 265                        & 2.85 & 205 & 2.20 \\
1 & 769 & 229 & 97 & 264 & 2.72 & 241 & 2.48 & 175 & 1.80 & 216 & 2.23 & 170 & 1.75 \\
2 & 725                        & 236                        & 93                        & 253                        & 2.72                        & 313                        & 3.37                        & 215                        & 2.31                        & 228                        & 2.45 & 197 & 2.12 \\
3 & 756                        & 227                        & 90                        & 281                        & 3.12                        & 232                        & 2.58                        & 206                        & 2.29                        & 258                        & 2.87 & 197 & 2.19 \\
4 & 850                        & 254                        & 99                        & 245                        & 2.47                        & 307                        & 3.10                        & 248                        & 2.51                        & 295                        & 2.98 & 226 & 2.28 \\
5 & 795                        & 258                        & 95                        & 285                        & 3.00                        & 309                        & 3.25                        & 227                        & 2.39                        & 268                        & 2.82 & 202 & 2.13 \\
6 & 864                        & 274                        & 110                       & 280                        & 2.55                        & 379                        & 3.45                        & 254                        & 2.31                        & 303                        & 2.75 & 230 & 2.09 \\
7 & 801                        & 245                        & 95                        & 298                        & 3.14                        & 331                        & 3.48                        & 240                        & 2.53                        & 260                        & 2.74 & 208 & 2.19 \\
8 & 893                        & 280                        & 109                       & 291                        & 2.67                        & 326                        & 2.99                        & 261                        & 2.39                        & 276                        & 2.53 & 232 & 2.13 \\
9 & 837                        & 238                        & 101                       & 282                        & 2.79                        & 324                        & 3.21                        & 251                        & 2.49                        & 277                        & 2.74 & 210 & 2.08 \\
\hline\hline
\multicolumn{4}{c|}{\textbf{geomean}}  & \textbf{} & \textbf{2.80} &              & \textbf{3.11}       & \textbf{}     & \textbf{2.31}    &               & \textbf{2.69}        & \textbf{}      & \textbf{2.11}   \end{tabular}%
}
\end{table*}

{\toqm} (\textbf{T}ime-\textbf{O}ptimal \textbf{Q}ubit \textbf{M}apping) \cite{Zhang+21-time} is an exact QCT algorithm for optimising the depth of the output circuit. Relaxation techniques are introduced to make {\toqm} scalable to 16 qubits and tens of thousands of gates. 
To evaluate the scalability of {\toqm} on large devices, we compared the performance of {\toqm}, \sabre, \sqgm\ on ten \quekno\ `20Q\_Tokyo\_depth' benchmarks \cite{li23quekno} on IBM Q Tokyo and 54Q Google Sycamore.
The results are presented in Tables~\ref{tab:toqm_on_tokyo} and ~\ref{tab:toqm_on_sycamore}. While Tables~\ref{tab:toqm_on_tokyo} confirms the efficacy of {\toqm} on the 20Q IBM Q Tokyo (1.88 vs. 2.39 and 1.91 for \sabre\ and \sqgm\ with repeat=100),  Table~\ref{tab:toqm_on_sycamore} shows that, running the same batch of benchmarks on 54Q Sycamore, {\toqm} does not perform  better than \sabre\ and \sqgm\ (2.80 vs. 2.69 and 2.11 for \sabre\ and \sqgm\ with repeat=100). We also tried to run the relaxed version of {\toqm} with \queko\ `54QBT\_5CYC\_QSE' (depth=5) circuits on 54Q Sycamore; however, it did not halt and return a result within 1,800 seconds. This indicates that, despite the relaxation techniques, {\toqm} is still not scalable to quantum devices with medium to large number of qubits.

\subsection{Compare with \nassc}\label{sec:nassc}
While most current QCT algorithms perform qubit routing independent of circuit optimisation,   \nassc\ (\textbf{N}ot \textbf{A}ll \textbf{S}WAPs have the \textbf{S}ame \textbf{C}ost) \cite{Liu+22_not_all_swap} combines qubit mapping with two optimisation techniques: 2-qubit block re-synthesis and commutation-based gate cancellation. It observes that (i) a SWAP gate could be merged into the previous 2-qubit block by re-synthesis to possibly reduce CX count by 3; (ii) a SWAP gate could be merged with a previous CX or SWAP gate to cancel two CXs. {\nassc} enriches \texttt{SabreSWAP} with the above optimisation techniques. 
 
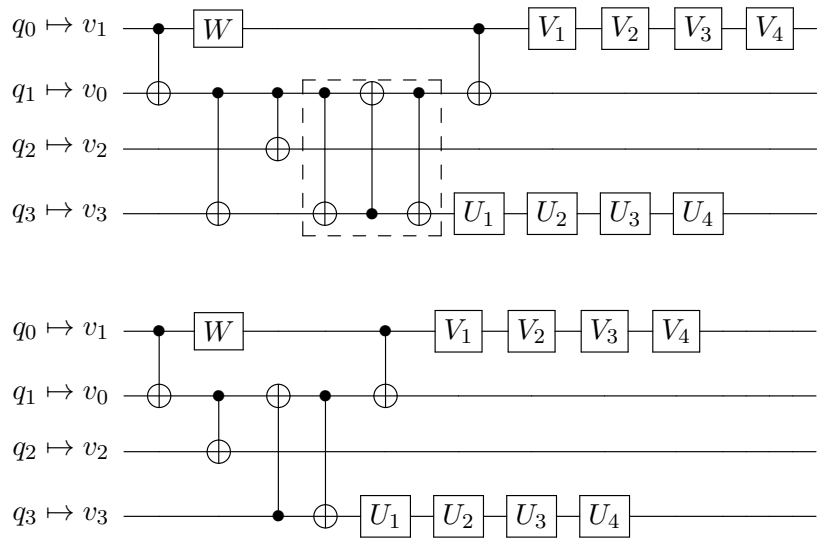
\begin{figure*}
\centering
\begin{tabular}{c}
\Qcircuit @C=0.8em @R=1.1em {
\lstick{q_{0}\mapsto v_{1}} &\ctrl{1} &\gate{W} &\qw      &\qw      &\qw       &\qw      &\ctrl{1}     &\gate{V_1} &\gate{V_2} &\gate{V_3} &\gate{V_4} &\qw \\
\lstick{q_{1}\mapsto v_{0}} &\targ    &\ctrl{2} &\ctrl{1} &\ctrl{2} &\targ     &\ctrl{2} & \targ       &\qw        &\qw        &\qw        &\qw        &\qw \\
\lstick{q_{2}\mapsto v_{2}} &\qw      &\qw      &\targ    &\qw      &\qw       &\qw      & \qw         &\qw        &\qw        &\qw        &\qw        &\qw \\
\lstick{q_{3}\mapsto v_{3}} &\qw      &\targ    &\qw      &\targ    &\ctrl{-2} &\targ    & \gate{U_1}  &\gate{U_2} &\gate{U_3} &\gate{U_4} &\qw        &\qw  
\gategroup{2}{5}{4}{7}{.7em}{--} %#1,3 rows #2,4 columns
}
\\
\quad
\\
\quad
\\
\Qcircuit @C=0.8em @R=1.1em {
\lstick{q_{0}\mapsto v_{1}} &\ctrl{1} &\gate{W} &\qw       &\qw      &\ctrl{1}     &\gate{V_1} &\gate{V_2} &\gate{V_3} &\gate{V_4} &\qw &\qw &\qw &\qw &\qw \\
\lstick{q_{1}\mapsto v_{0}} &\targ    &\ctrl{1} &\targ     &\ctrl{2} & \targ       &\qw        &\qw        &\qw        &\qw        &\qw &\qw &\qw &\qw &\qw \\
\lstick{q_{2}\mapsto v_{2}} &\qw      &\targ    &\qw       &\qw      & \qw         &\qw        &\qw        &\qw        &\qw        &\qw &\qw &\qw &\qw &\qw \\
\lstick{q_{3}\mapsto v_{3}} &\qw      &\qw      &\ctrl{-2} &\targ    & \gate{U_1}  &\gate{U_2} &\gate{U_3} &\gate{U_4} &\qw        &\qw &\qw &\qw &\qw &\qw \\
}
%\caption{After cancelling $CX(v_0,v_3)$ gates (depth=9).}
%\end{subfigure}
\end{tabular}
\caption{Another better transformation with depth 11 (top). After applying commutative gate cancellation (cancelling the first two $CX(v_0,v_3)$ in the top circuit) the circuit depth is further reduced to 9 (bottom) }\label{fig:betterrun2}
\end{figure*}

Continuing with our running example, Fig.~\ref{fig:betterrun2} shows another better transformation, which inserts $\swap{v_0}{v_3}$ instead of $\swap{v_0}{v_1}$ as in Fig.~\ref{fig:betterrun}. Note that the first CX of $\swap{v_0}{v_3}$ is commutable with the CX before it. If we commute and cancel it with the first $CX(v_0,v_3)$, we shall have a circuit with depth 9. One merit of \nassc\ is its ability to identify this possibility and, then, use the right SWAP decomposition in the post-routing optimisation.

 {\nassc} has the same time complexity as \sabre\ and experiments show that it can reduce by 21.30\% and 7.61\%, on average, the number of CX gates and circuit depth, respectively, when compared with \sabre. As {\nassc} applies the Qiskit \texttt{Commutative Gate Cancellation} pass, it is not clear how much of this reduction is contributed solely by \nassc's routing pass. 
 
While there are eight combinations of the three optimisations (2-qubit gate block re-synthesis and two commutation-based optimisations), enabling all three optimisations produces a similar performance to the best combination \cite{Liu+22_not_all_swap}. Hence, we assume all three optimisations as well as the following are enforced in \nassc:\footnote{Unitary synthesis was initially included for both pre- and post-optimisations, but was removed in our experiments as it was found that it led to significant increase in transformed circuit depth and running time.} 
\begin{itemize}
    \item \textit{pre-routing optimisation}: involves combining chains of single-qubit gates into a single gate (single-qubit decomposition optimisation), 
    \item \textit{commutative gate cancellation (CC)}: involves iteratively cancelling gates based on commutation rules until no changes in depth are observed, and 
    \item \textit{post-routing optimisation}: involves both single-qubit decomposition optimisation and commutative gate cancellation. 
\end{itemize}
For a fair comparison with \nassc, we also enhance the base \sabre\ and \sqgm\ algorithms with a post-routing \textit{commutative gate cancellation (CC)}.

We then compare the performance of \sabre, \sqgm, and {\nassc} on selected \texttt{QUEKO} and MQTBench library circuits \cite{quetschlich2022mqtbench} on  the 54Q Google Sycamore. To ensure fairness, the same initial mapping generated by Qiskit's \textsf{SabreLayout} was used across the algorithms for each repeat. 

\begin{table}[h]
\centering
\caption{Comparison of \sabre\ and \sqgm\ with \nassc\ on 10 \queko\ `54QBT\_25CYC\_QSE' benchmarks. }\label{tab:sum_nassc_quekno}
\resizebox{0.48\textwidth}{!}{%
\begin{tabular}{c|ccccc}
algorithm & \sabre & \sabre+CC & \sqgm & \sqgm+CC &\nassc \\ \hline
avg. depth ratio        & 5.18       & 4.00          & 3.34          & 2.88          & 3.96          \\
ratio with SABRE & - & \textbf{0.77} & \textbf{0.64} & \textbf{0.56} & \textbf{0.76}\\
\bottomrule
\end{tabular}
}
\begin{tablenotes}
\item [a] Note: each entry in row `avg. depth ratio' is the geomean of best depth ratios obtained from 5 repeats for ten \queko\ circuits.
\end{tablenotes}
\end{table}

The first set of tests comprised the aforementioned \queko\ `54QBT\_25CYC\_QSE' circuits.  Considering the base algorithms alone, \nassc\ and \sqgm\  yielded much lower average depth ratios than \sabre\ (3.96 and 3.34  vs. 5.18). When post-routing CC optimisation is adopted, the ratios for \sabre+CC and \sqgm+CC decrease sharply to 4.00 and 2.88. 
While \nassc\ remains slightly better than \sabre+CC, \sqgm+CC beats \nassc\ by $1-2.88/3.96=27\%$.  The unexpected sharp decrease obtained with CC is due to the presence of Pauli $X$ gates as the sole kind of single-qubit gates in \queko\ circuits, thus enabling CC to cancel many gates. 

The second set of tests comprised 117 real circuits from the MQT Bench library, with qubit number ranging from 50 to 54 qubits (cf.~Sec.~\ref{sec:bench}). For these circuits, the number of repeats was 5. Table~\ref{tab:mqtbench} shows a part of the results for those circuits with 53 qubits. We observe from the table that, firstly, \nassc\ and \sqgm\ reduce the depth by 17\% and 27\% against \sabre. Secondly, enhanced with post-routing CC, the depth of \sabre\ is reduced by 7\%, which is quite modest compared with the result for \queko\ benchmarks shown in Table~\ref{tab:sum_nassc_quekno}; this, however, should be more representative of the general case. Lastly, \sqgm+CC obtains the best average result, which shows that enhancing with CC will further boost the depth optimality of \sqgm.

\section{Conclusion}
\label{sec:conclusion}
While we have recently witnessed significant advances in building large quantum computers, qubit coherence time remains a strict restriction to NISQ devices. In this sense, it seems more important for qubit mapping to minimise the depth overhead. In this paper, we demonstrated that a simple SWAP gate insertion may double the depth of a circuit, from which we proposed a simple yet general method that takes into consideration the impact of single-qubit gates on circuit depth. The effectiveness of our method was demonstrated by embedding it in \sabre, and our experiments on three architectures and extensive benchmarks confirmed that the new algorithm, called \sqgm, is able to significantly reduce the depth overhead of circuit transformation when compared with \sabre.  Compared with the relaxed version of {\toqm} \cite{Zhang+21-time},  a time-optimal algorithm, our method is scalable to large devices. In addition, its performance can be further enhanced with post-routing commutative gate cancellation.

In real quantum devices, the reliability of qubits and qubits connectivity strengths vary spatially and temporally. Further work is required to extend our method to QCT algorithms which respect these hardware characteristics \cite{Murali+19,TannuQ19} or those that  mitigate crosstalk noises \cite{Xie21dac_commutativity}. 

\section*{Acknowledgements}
This work was partially supported by the Australian Research Council (Grant No.: DP220102059) and the National Science Foundation of China (Grant No.: 12071271). Ky Dan and Zachary were also partially supported by Sydney Quantum Academy (SQA) under two SQA Undergraduate Grants.
\bibliographystyle{IEEEtran}
\bibliography{qct-lsj}
\end{document}